\documentclass[prb,amsmath,amssymb,showpacs,preprint]{revtex4}
\usepackage{graphicx}
\def\be{\begin{equation}}
\def\ee{\end{equation}}

\begin{document}
\title{Influence of the boundary conditions on the current flow pattern along a superconducting wire}
\author{Jorge Berger}
\affiliation{Department of Physics, Ort Braude College, 21982 Karmiel, Israel} 
\email{jorge.berger@braude.ac.il}
\begin{abstract}
We study the patterns at which the current flow stabilizes in a 1D superconducting wire, for various experimentally reasonable boundary conditions, for small fixed current densities and temperatures close to $T_c$. We pay special attention to the possible existence of a stationary regime. If the contacts are superconducting, truly stationary or normal regimes do not exist, but can be approached as a limit. In the case of weak superconducting contacts, a rich phase diagram is found, with several periodic regimes that involve two phase slip centers. For some of these regimes, the density of Cooper pairs does not have mirror symmetry. If the contacts are normal, the stationary regime is possible.
\end{abstract}
\maketitle

\section{Introduction: The questions we intend to answer}
When current is driven along a superconducting wire, it can either flow as normal current, as supercurrent, or as a combination of both. The various patterns that may be obtained in space and time were reviewed long ago,\cite{Ivlev,Tidecks} and a new approach\cite{koby} has been raised in recent years. The results obtained for 1D wires have been extended to the case of stripes.\cite{Lydia,Berdiy,Almog}

One of the possible patterns is periodic in time, and a salient feature of the periodic regime is the appearance of phase slips that occur when the superconducting order parameter vanishes at some point. Phase slips occur at definite positions, called phase slip centers (PSC). While early studies (e.g.\ Refs.\ \onlinecite{Ivlev}--\onlinecite{Tidecks} and references therein) were mainly interested in wires of effectively infinite length, so that they extended over many PSC and were insensitive to the boundary conditions, Ref.~\onlinecite{koby} focuses on a parameter region in which there are just a few PSC, if any. 
This is the region that most neatly exhibits the qualitative features of the current pattern and, with present microfabrication techniques, it becomes an experimentally relevant region.

\begin{figure}
\scalebox{0.85}{\includegraphics{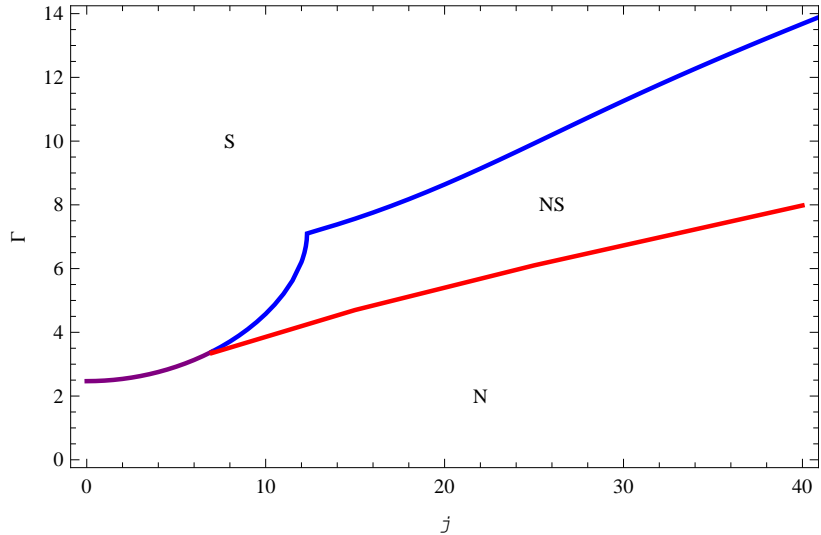}}
\caption{\label{r0}(Color online) Phase diagram in the current density-temperature plane for the current flow regimes, if the order parameter vanishes at the extremes of the wire and $u=5.79$. ``S" denotes stationary regime, ``N" denotes exclusively normal current, and ``NS" denotes a region where both regimes are possible. The blue line is the stability limit when either the current or the temperature decreases (i.e., the normal state is stable up to this line), the red line is the stability limit when they increase (i.e., the stationary state is stable down to this line), and the purple line is the stability limit in both directions. For a wire of length $2L$, the values of $\Gamma $ in this graph have to be multiplied by $(L/\xi (0))^{-2}$ and those of $j$  by $(L/\xi (0))^{-3}$.}
\end{figure}

Figure \ref{r0} shows the phase diagram for the flow patterns found in Ref.~\onlinecite{koby} for small currents and for temperatures close to $T_c$. Below the pertinent curve the sample is in the normal state (N) and all the current is normal, whereas above the curve there is a stationary regime (S), in which normal current and supercurrent are both present, both are functions of position, but none of them depends on time. The stationary regime was previously found in Ref.~\onlinecite{Bara}. The third possibility found in Ref.~\onlinecite{koby} is the periodic regime, in which the normal current and the supercurrent are periodic functions of time; however, following a similar approach to Ref.~\onlinecite{koby}, it was found\cite{shim} that for realistic material parameters the periodic regime is strongly disfavored in comparison to the stationary state.

We must therefore confront an intriguing situation: on the one hand, phase slips are described in textbooks and their presence has been confirmed experimentally, whereas the stationary possibility is practically ignored in the literature; on the other hand, Refs. \onlinecite{koby}, \onlinecite{Bara} and \onlinecite{shim} suggest that the stationary regime prevails. One reason for this disparity is the small effective length of the wires considered in Refs. \onlinecite{koby}, \onlinecite{Bara} and \onlinecite{shim}. We argue that the other reason is that these studies assume that the order parameter vanishes at the boundaries.

The order parameter does vanish at the boundaries if the ``banks" at which the current is fed contain ferromagnetic impurities. In the present study we will compare this situation with other possibilities. The banks will be either superconducting or normal. We may then consider the limiting cases of very weak superconductors or of normal metals with very short de Gennes length, and examine in what sense the stationary regime is approached.

\section{Model \label{Mod}}
Following Refs.\ \onlinecite{koby} and \onlinecite{Bara}, we use a minimal model. The wire will be perfectly 1D and uniform, we will assume electroneutrality, and superconductivity will be described by the time-dependent Ginzburg-Landau model (TDGL). We choose a gauge with no vector potential, write $\Gamma =1-T/T_c$  and  denote by $\varphi$ the electrochemical potential. The unit of length will be denoted by $x_0$ , by $t_0$ the unit of time, by $\varphi_0$ the unit of voltage, and by $j_0$ the unit of the current density. As in Ref.~\onlinecite{shim}, we take
\begin{equation}
x_0=\xi(0)\;,\;\;t_0=\frac{\pi\hbar }{8k_BT_c}\;,\;\;\varphi_0=\frac{4k_BT_c}{\pi e}\;,\;\;
j_0=\frac{4\sigma  k_BT_c}{\pi e\xi (0)}\;.
\label{units}
\end{equation}
Here $\xi (0)$ is the coherence length at $T=0$, $k_B$ is Boltzmann's constant, $e$ is the electron charge, and $\sigma $ is the normal conductivity.

With this notation, the TDGL equation and Ohm's law read 
\begin{equation}
\psi _t  = \psi _{xx}  + \left( {\Gamma  -  \left| \psi  \right|^2  - i\varphi } \right)\psi \;
\label{tdgl}
\end{equation}
and
\begin{equation}
\varphi _x  = u\, {\rm Im}\left( { \psi _x \psi ^* } \right) - j\;.
\label{Ohm}
\end{equation}
Here $\psi$ is the order parameter, with normalization imposed by Eq.~(\ref{tdgl}), the subscripts denote partial differentiation with respect to the time $t$ and the arc length $x$ along the wire, and $u$ is the ratio between the relaxation times of $\psi$ and $j$.\cite{Kopnin} 

Our model neglects Joule heating and thermal fluctuations. Joule heating can be neglected provided that the current density is sufficiently small and/or the wire is sufficiently thin and in good thermal contact with a heat bath. Thermal fluctuations are expected to be important near stability limits.

The wire is assumed to extend along $-L\le x\le L$. 
Equations (\ref{tdgl}) and (\ref{Ohm}) are invariant under the transformation $x\rightarrow Lx$, $t\rightarrow L^2t$, $\psi\rightarrow L^{-1}\psi$, $\varphi \longrightarrow L^{-2}\varphi $, $\Gamma \rightarrow L^{-2}\Gamma $ and $j\rightarrow L^{-3}j$, since each of the terms in Eqs.\ (\ref{tdgl}) and (\ref{Ohm}) is multiplied by $L^{-3}$. Therefore, if also the boundary conditions are invariant under this transformation, we can limit our study to a single value of $L$, and the solutions for any other value can be obtained by scaling. Following Ref.~\onlinecite{koby}, we will set $L=1$.

\section{Banks of same material as the wire}
The most common experimental situation is that the banks and the wire are carved from the same layer. The banks are located at the extremes of the wire, $x=\pm L$, and are much wider than the wire, so that the current density in them is negligible. The banks can therefore be treated as being in equilibrium, and we obtain from Eq.~(\ref{tdgl}) that $|\psi (\pm L)|=\Gamma^{1/2}$. The phase of the order parameter has to obey the Josephson relation and we therefore require the boundary conditions
\be
\psi (\pm L,t)=\Gamma^{1/2}\exp \left[ -i \int_0^t\varphi (\pm L,t')dt'\right] \;.
\label{bcSC}
\ee
The electrochemical potential $\varphi $ is obtained from Eq.~(\ref{Ohm}), in which we will set $u=5.79$, as appropriate for a superconducting material with paramagnetic impurities in the dirty limit.

Equations \ref{tdgl}--\ref{bcSC} were solved numerically; numerical details are provided in the Appendix. Figure \ref{r1} shows the phase diagram that we encounter in the present case. In contrast to Fig.~\ref{r0}, the possible regimes are either purely superconducting (SC) or periodic.
It is clear that a strictly normal regime is incompatible with the condition (\ref{bcSC}), since the order parameter cannot vanish close to $x=\pm L$. Also the stationary regime is implausible: normal currents would imply dissipation, so that $\varphi (L)<\varphi (-L)$. In this case condition (\ref{bcSC}) implies winding that increases with time, and can only be released if the order parameter vanishes at some point, so that phase slips are expected.
\begin{figure}
\scalebox{0.85}{\includegraphics{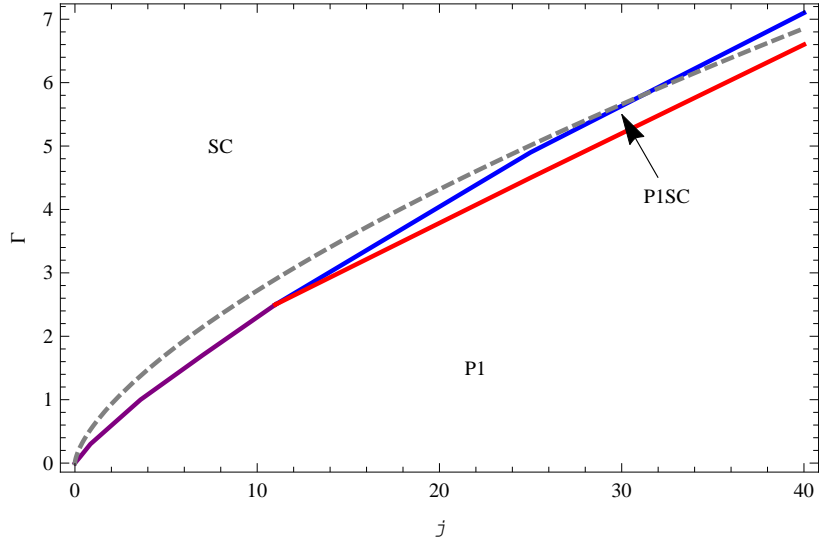}}
\caption{\label{r1}(Color online) Phase diagram if the contacts to the wire are made of the same superconducting material as the wire. ``SC" stands for exclusively superconducting current, ``P1" stands for periodic regime with one phase slip center at the middle of the wire, and ``P1SC" denotes a region where both regimes are possible. The blue line is the stability limit when either the current or the temperature decreases, the red line is the stability limit when they increase, and the purple line is the stability limit in both directions. The dashed grey curve corresponds to the critical current density $j_c=2u(\Gamma /3)^{3/2}$, that would be obtained in a very long wire in which the boundaries become irrelevant.}
\end{figure}

It should be mentioned that although in the periodic regime there is a voltage drop along the wire, this problem is not equivalent to the case of fixed applied voltage studied in, e.g., Refs.\ \onlinecite{S} and \onlinecite{Lydia1}. In the present case $\varphi (-L)-\varphi (L)$ is periodic, rather than constant in time.

If there is no normal current and if the order parameter has uniform size, then Eqs.\ (\ref{tdgl}) and (\ref{Ohm}) imply that the maximum current density is $j_c=2u(\Gamma /3)^{3/2}$. $j_c$ is called the `critical current density', and it is usually stated that the wire is in the normal state for $j>j_c$. 
In Fig.~\ref{r1} we see that the condition $|\psi (\pm L)|=\Gamma^{1/2}$ favors the superconducting regime for small currents, but inhibits it for large currents.
\begin{figure}
\scalebox{0.7}{\includegraphics{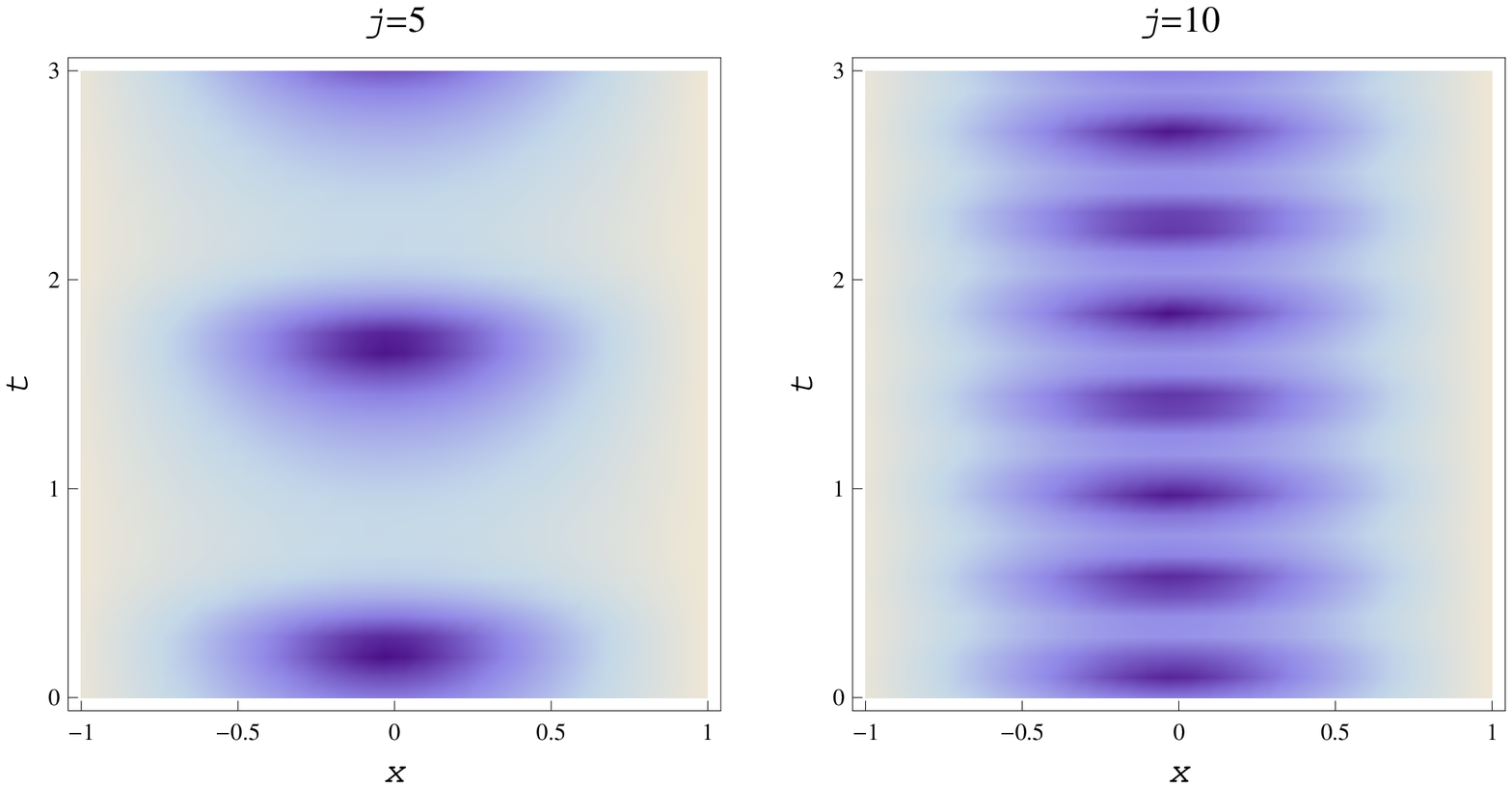}}
\caption{\label{P1normal}(Color online) Density plot of the absolute value of the order parameter, $|\psi (x,t)|$, for current densities above the ``critical current density." In these plots we took $\Gamma =1$, so that they both correspond to the periodic regime. $|\psi |=1$ at the boundaries, and vanishes at the phase slip points. $j_c=2.23$.}
\end{figure}

Since the wire cannot be strictly normal, we investigate in what sense the normal regime is approached when $j$ rises above $j_c$. Figure \ref{P1normal} shows the size of the order parameter $|\psi (x,t)|$ for two current densities above $j_c$. We see that the periodic regime persists, but for larger currents phase slips are more frequent, and therefore the order parameter has less time to recover and remains small.

\section{Banks of a superconducting material that is weaker than that of the wire}
We consider now the case that the wire and the banks are made of different superconducting materials, so that $|\psi (\pm L,t)|\neq \Gamma^{1/2}$. We write $|\psi (\pm L,t)|=r\Gamma^{1/2}$ and replace the boundary condition (\ref{bcSC}) with
\be
\psi (\pm L,t)=r\Gamma^{1/2}\exp \left[ -i \int_0^t\varphi (\pm L,t')dt'\right] \;.
\label{rSC}
\ee
We will especially be interested in the case that $r$ is significantly less than 1.

\begin{figure}
\scalebox{0.85}{\includegraphics{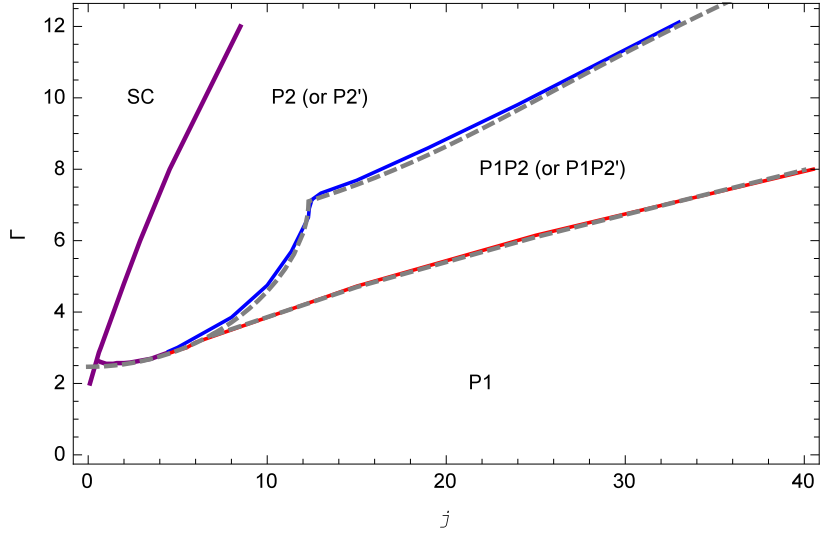}}
\caption{\label{r05}(Color online) Phase diagram if the order parameter at the contacts is smaller by a factor 0.05 than what it would be if the banks and the wire were of the same material. ``P2" (or ``P2$'$") stands for periodic regime with two phase slip centers, ``SC" and ``P1" have the same meaning as in Fig.\ \ref{r1}, and ``P1P2" (or ``P1P2$'$") denotes a region where one or two PSC are possible.
The blue line is the stability limit when either the current or the temperature decreases, the red line is the stability limit when they increase, and the purple lines are the stability limit in both directions. 
The dashed grey curves are the stability lines for $r=0$, presented in Fig.\ \ref{r0}.}
\end{figure}

In this case we find a surprisingly rich phase diagram. Figure \ref{r05} shows the phase diagram obtained for $r=0.05$. In addition to the regimes SC and P1, found for $r=1$, we find additional regimes that mediate between them. These regimes are periodic with two PSC at symmetric positions with respect to the middle of the wire. Sometimes we found that both phase slips occur simultaneously and we denote this regime by P2. We also found cases in which the two phase slips are separated in time by half a period, and denote this regime by P2$'$.
The stabilization time required to pass between the situations P2 and P2$'$ is very long (several times  $10^3\,t_0$). We did not investigate the stability boundaries between P2 and P2$'$ for $r=0.05$, but this will be done for $r=0.2$.

The upper stability boundary of P1 almost coincides with the upper stability boundary of the normal regime in the case $\psi (\pm L)=0$, and the lower boundary of P2 almost coincides with the lower boundary of the stationary regime. 

It should be born in mind that, for given materials, $r$ is a function of temperature. Therefore, the phase diagrams for fixed $r$ should not be directly interpreted as phase diagrams in the current density-temperature plane.

\begin{figure}
\scalebox{0.85}{\includegraphics{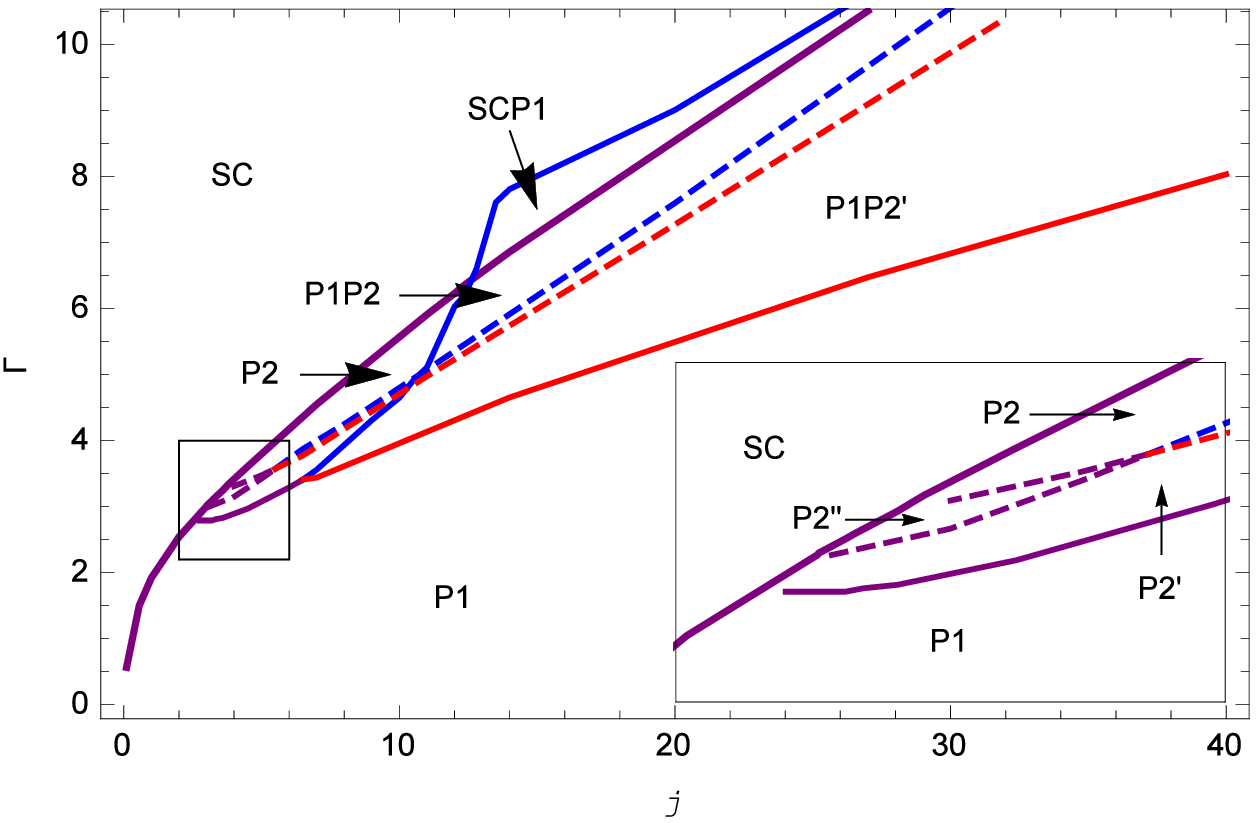}}
\caption{\label{r2}(Color online) Phase diagram for $r=0.2$. The colors of the stability boundaries, as well as ``P2", ``SC", ``P1" and ``P1P2", have the same meaning as in Fig.~\ref{r05} and, likewise, SCP1 and P1P2$'$ are regions where two regimes are possible. The stability boundaries that involve two regimes that both have two PSC are depicted by dashed lines. The framed region $2\le j\le 6$ and $2.2\le\Gamma\le 4$ is shown in an enlarged scale in the inset. Between the red and the blue dashed lines, P2, P2$'$ and P1 are possible.}
\end{figure}

As $r$ increases, the region occupied by the purely superconducting regime increases, and the topology of the phase diagram can change. Figure \ref{r2} shows the phase diagram for $r=0.2$. We note that in this case there is a region where SC and P1 are both possible and that the stability boundaries of P2 and P2$'$ have been investigated separately.
                          
\begin{figure}
\scalebox{0.7}{\includegraphics{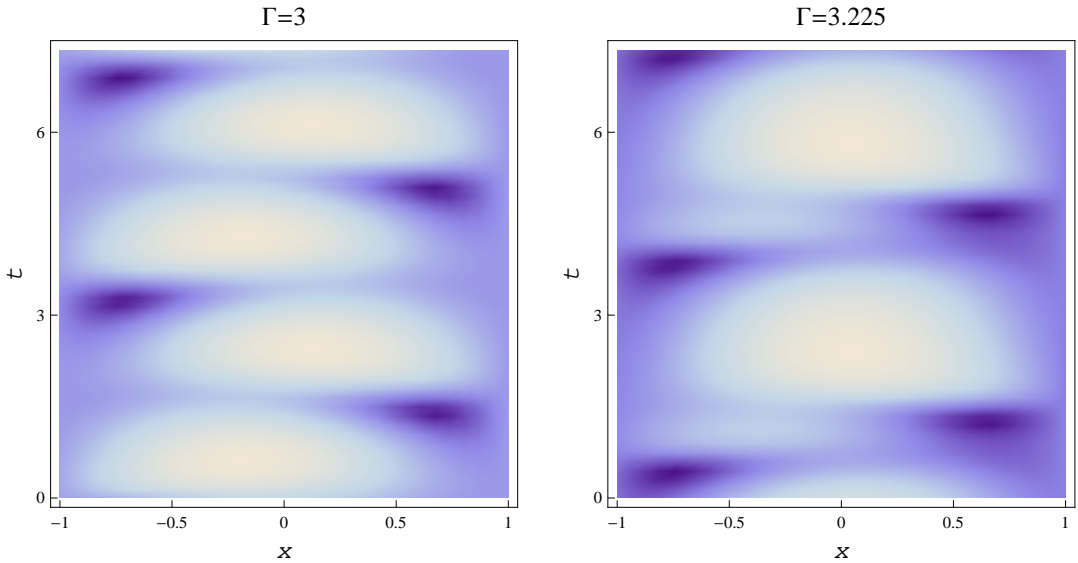}}
\caption{\label{asymPSC}(Color online) Density plots of $|\psi (x,t)|$ for $r=0.2$ and $j=4$. Darker purplish color means smaller $|\psi |$. The case $\Gamma =3$ is in the P2$'$ regime and the case $\Gamma =3.225$ is in the P2$''$ regime.}
\end{figure}

For $j\alt 5.5$, the transition between P2 and P2$'$ in Fig.~\ref{r2} is continuous. In this range P2 and P2$'$ are mediated by a regime with two PSC, that we denote by P2$''$, in which the time lag changes gradually with decreasing $\Gamma $ from zero to half a period. Figure \ref{asymPSC} shows density plots of $|\psi (x,t)|$ for $j=4$, for $\Gamma =3$ (P2$'$ regime) and for $\Gamma =3.225$ (P2$''$ regime). We are not aware of previous reports of situations in a uniform superconducting wire such that the order parameter obeys $|\psi (-L,t)|=|\psi (L,t)|$, but stabilizes in a regime in which $|\psi (-x,t)|\neq |\psi (x,t)|$.

It is hard to judge numerically whether situations as the case $\Gamma =3.225$ in Fig.~\ref{asymPSC} truly correspond to a distinct P2$''$ regime, or are rather encountered due to insufficient stabilization time. We believe they correspond to a distinct regime, because starting from different initial order parameters we obtain the same time lag between consecutive phase slips.

\begin{figure}
\scalebox{0.85}{\includegraphics{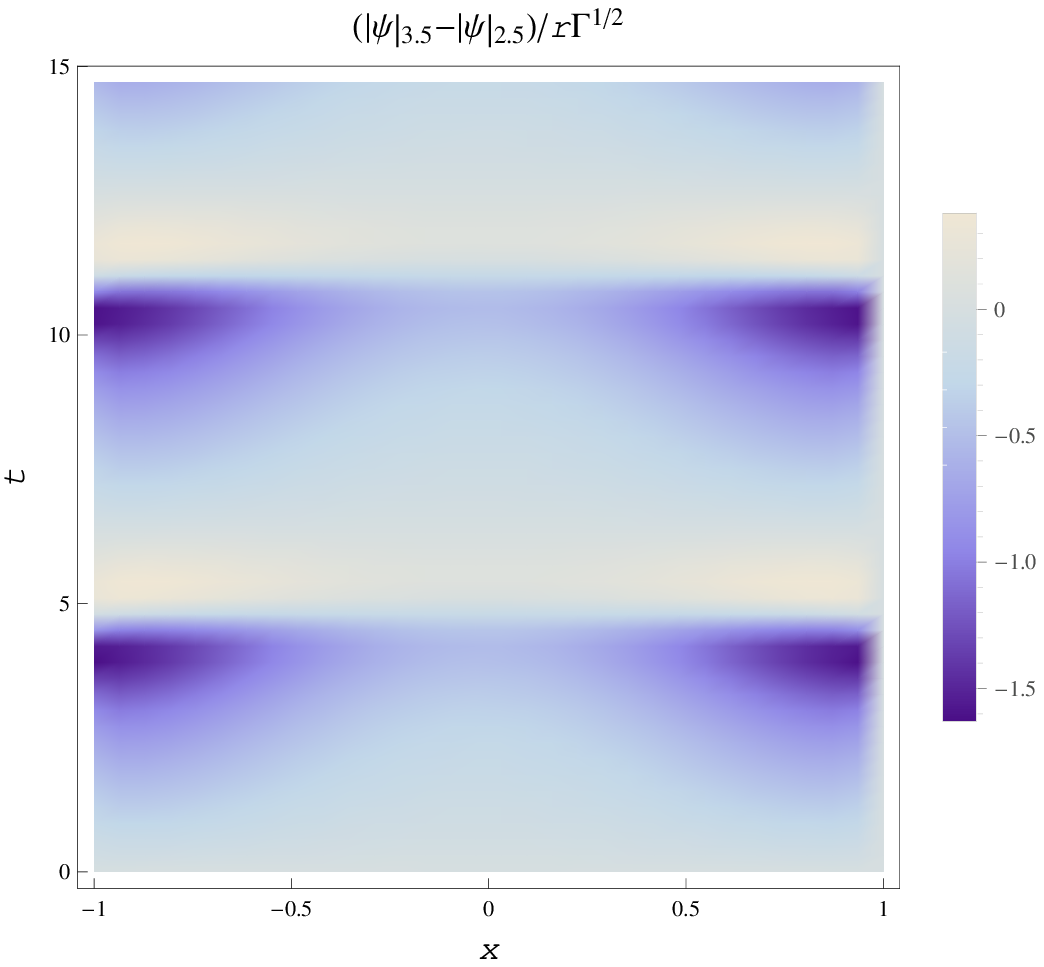}}
\caption{\label{hefreshP2}(Color online) Difference between the size of the order parameter in the P2 regime and in the SC regime. In both cases we have taken $\Gamma =6$ and $r=0.05$. $|\psi |_{3.5}$ denotes $|\psi (x,t)|$ for $j=3.5$ (in this case the system is in the P2 regime) and $|\psi |_{2.5}$ denotes $|\psi (x,t)|$ for $j=2.5$ (in this case the system is in the SC regime).}
\end{figure}

\begin{figure}
\scalebox{0.85}{\includegraphics{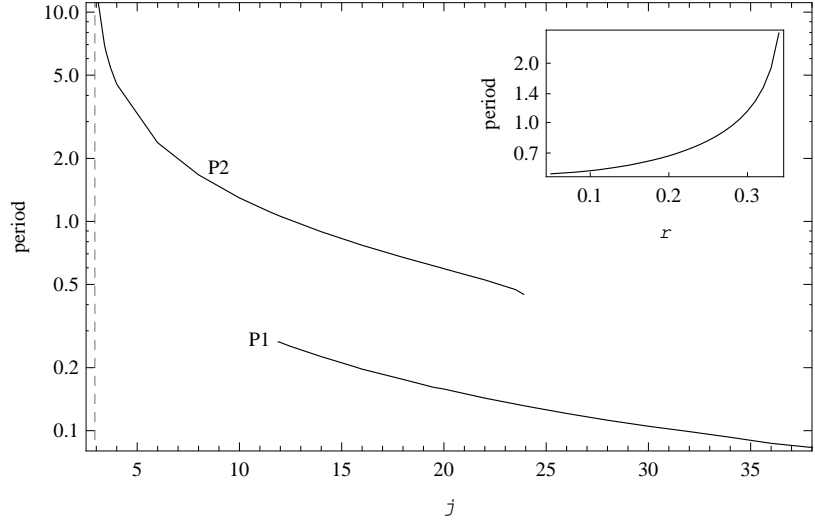}}
\caption{\label{periodofj} Duration of a period as a function of the current density for the P2 regime (upper curve) and for the P1 regime (lower curve), for $\Gamma =6$ and $r=0.05$. The curves end where these regimes stop to exist. The grey dashed line indicates the transition between P2 and SC. The $y$-axis is logarithmic. Inset: Period in the P2 regime as a function of $r$, for $\Gamma =7$ and $j=24$; for $r>0.34$, only supercurrent is present. }
\end{figure}

Since Figs.\ \ref{r05} and \ref{r2} show no hysteresis in the transition between SC and P2, we expect a continuous transition. At first sight this seems impossible, since in P2 the order parameter approaches zero at the phase slips, whereas in SC it does not. Indeed, near the phase slips a small change in the current density that leads to a transition can result in large differences in the order parameters, as illustrated in Fig.~\ref{hefreshP2}. However, as shown in Fig.~\ref{periodofj}, the time between consecutive phase slips diverges when the SC regime is approached, so that P2 approaches SC almost everywhere in the $xt$ plane. The inset in the figure shows that the same effect occurs when the P2$\rightarrow $SC transition is due to increase of $r$.
\begin{figure}
\scalebox{0.85}{\includegraphics{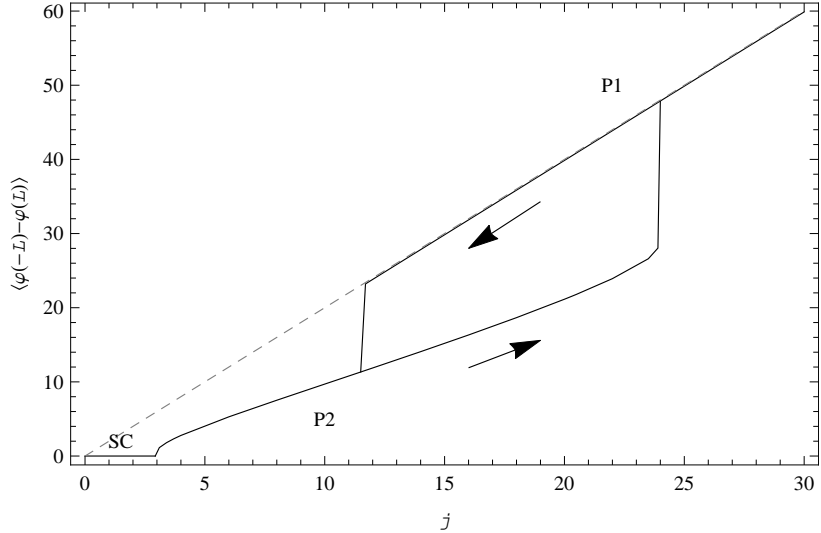}}
\caption{\label{Vofj} Time average of the voltage drop along the wire, as a funtion of the current density, for $\Gamma =6$ and $r=0.05$.  The arrows indicate the direction in which the current is varied. The grey dashed line shows the voltage that would be obtained if the wire were in the normal state. }
\end{figure}

An experimentally accessible quantity is the dc component of the voltage drop along the wire. Figure \ref{Vofj} shows this quantity for the same parameters as in Fig.~\ref{periodofj}, which involve a hysteresis region.

For the parameters chosen in Figs. \ref{periodofj} and \ref{Vofj} the position of the PSC in the P2 regime are almost independent of $j$. Also in the P2$'$ regime these positions are almost fixed in most of the stability range, but in the case of a continuous transition to P1, very near the transition, the PSC migrate to the center of the wire, and thus the P1 regime is obtained and the period is halved.

\begin{figure}
\scalebox{0.85}{\includegraphics{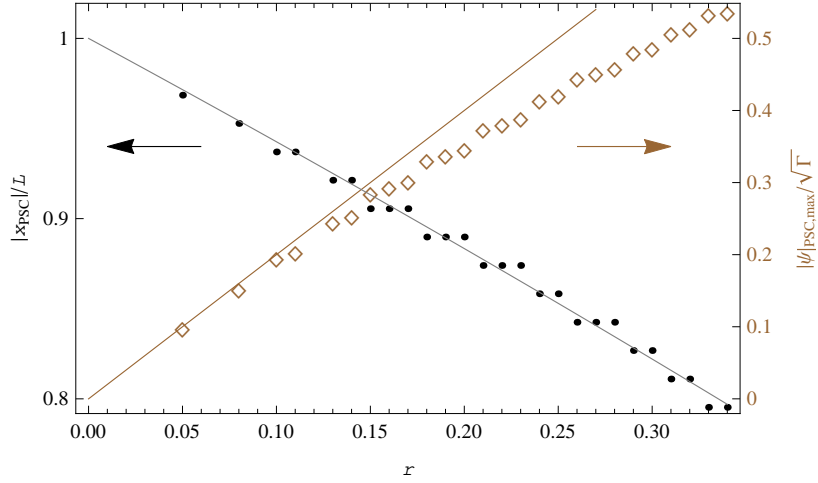}}
\caption{\label{PSCofr}(Color online) The black dots (left $y$-axis) show the distance beween the phase slip centers and the middle of the wire in the P2 regime as a function of $r$, for $\Gamma =7$ and $j=24$. The brown squares (right $y$-axis) show the maximum size of the order parameter at these phase slip centers. The grey line is a guide for the eye and the brown straight line has slope 2. The staircase appearance is due to the finite discretization of position.}
\end{figure}

We are now in a position to discuss in what sense the P2 regime approaches the stationary regime in the limit $r\rightarrow 0$. We first note from Fig.~\ref{PSCofr} that in this limit the PSC approach the extremes $x=\pm L$ of the wire, and that at these PSC the order parameter $\psi_{\rm PSC}$ is bound to the size $|\psi_{\rm PSC}|\leq 2r\Gamma^{1/2}$. Therefore $\psi (\pm L)\rightarrow 0$ for $r\rightarrow 0$, as expected, and there are no phase slips in the interior of the wire. 

Next, we recall that a peculiar measurable property of the stationary regime is that the potential felt by Cooper pairs, $-\varphi_0t_0\partial \arg\psi /\partial t$ (that can be probed by means of SIS junctions, as in Ref.~\onlinecite{DJ}) is independent of position. Quite generally, we can conclude that if $x=x_1$ and $x=x_2$ are two points of the wire such that $\psi (x_1,t)$ and $\psi (x_2,t)$ are periodic and there is no PSC in the segment $x_1\le x\le x_2$, then the time averages of $\partial \arg\psi /\partial t$ at $x=x_1$ and at $x=x_2$  have to be equal. Otherwise, $|\arg\psi (x_1)-\arg\psi (x_2)|$ would endlessly increase, the order parameter would become increasingly wound up, and this would force phase slips in the segment $x_1\le x\le x_2$. Summing up, we conclude that in the P2 regime the time average of $\partial \arg\psi /\partial t$ is uniform in the segment between the PSC, and in the limit $r\rightarrow 0$ this property applies to the entire wire.

\begin{figure}
\scalebox{0.85}{\includegraphics{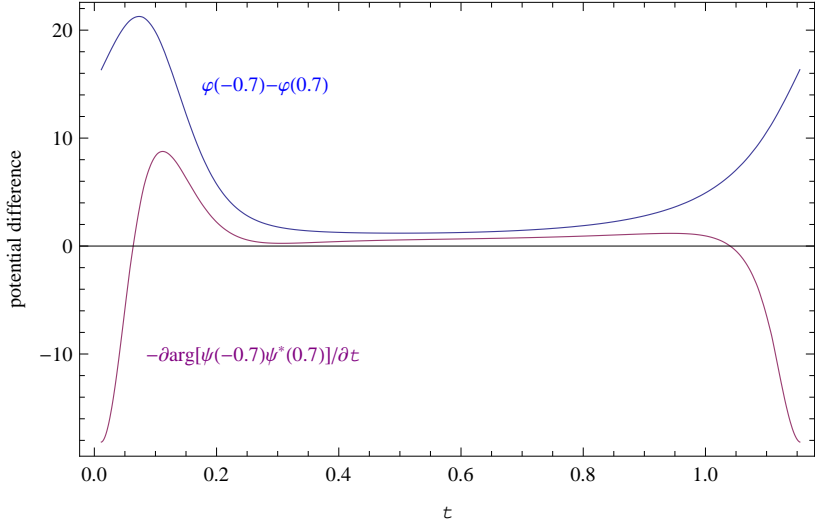}}
\caption{\label{vn7vc7}(Color online) Potential differences between the points $-0.7L$ and $0.7L$ felt by normal electrons and by Cooper pairs, as functions of time, for a situation in the P2 regime. The graph extends over one period, and the origin of time was set when a phase slip occurs. The parameters used were $\Gamma =7$, $j=24$, and $r=0.3$. For these parameters the PSC are at $\pm 0.83L$. For a wire of length $2L$, these parameters correspond to a temperature $T_c[1-7(\xi (0)/L)^2]$, current density $j_0(\xi (0)/L)^3$, the potential differences equal the value in the graph times $\varphi_0 (\xi (0)/L)^2$, and time equals the value in the graph multiplied by $t_0(L/\xi (0))^2$.}
\end{figure}
 
In contrast to the stationary regime, in the case of P2 the potential felt by Cooper pairs is not independent of position for arbitrary times in the segment between the PSC. Figure \ref{vn7vc7} shows the potentials felt by Cooper pairs and by normal electrons between two given points, as functions of time, during a period. The normal potential peaks shortly after the phase slips.

\section{Banks of a normal metal}
In the spirit of a minimal model description, we adopt the de Gennes boundary condition\cite{dG}
\be
\psi_x(\pm L)=\mp\psi (\pm L)/b \;,
\label{DG}
\ee
where $b$ is a length that represents how far Cooper pairs can survive inside the normal metal. The limit $b\rightarrow \infty $ corresponds to an insulator and $b\rightarrow 0$ corresponds to a metal with magnetic impurities. Condition (\ref{DG}) can be justified microscopically for static situations, and we may expect that it is still qualitatively correct for small current densities. Since condition (\ref{DG}) implies that there is no supercurrent at the contacts, it may also be an appropriate description of the case in which quasiparticles are injected and withdrawn from the wire by means of NIS junctions. The scaling with length introduced at the end of Sec.~\ref{Mod} has to be supplemented with $b\rightarrow Lb$.

Within a numerical scheme in which position along the wire is replaced by a computational grid and $\psi_x(\pm L)$ are approximated by finite differences, condition (\ref{DG}) becomes 
$\psi (\pm L)=b\psi (\pm L\mp\Delta x)/(b+\Delta x)$, where $\Delta x$ is the length of a segment in the grid.

Unlike the previous sections, and since the points $x=\pm L^\mp$ are not in equilibrium, the rate of change of the phases at these points is not dictated by the potential, so that it is possible to have $\varphi (-L)\neq\varphi (L)$ without phase slips, and the stationary regime is not ruled out.

The stability boundary of the normal regime can be found as in Ref.~\onlinecite{koby}, since the evolution equations and the boundary conditions are PT-symmetric,\cite{koby} and all the theorems shown in Ref.~\onlinecite{Zum} apply here as well. The normal regime is unstable if Eqs.\ (\ref{tdgl}), (\ref{Ohm}) and (\ref{DG}) have a solution of the form $\psi (x,t)=f(x)e^{(\Gamma -\gamma )t}$ with Re$(\gamma )<\Gamma $. At the bifurcation from N the nonlinear terms in Eqs. (\ref{tdgl}) are (\ref{Ohm}) negligible, and we are left with the spectral problem
\be
f_{xx}+ixjf=-\gamma f, \;\;\;f_x(\pm L)=\mp f (\pm L)/b \;,
\ee
where $\gamma $ is the eigenvalue with the smallest real part.

\begin{figure}
\scalebox{0.85}{\includegraphics{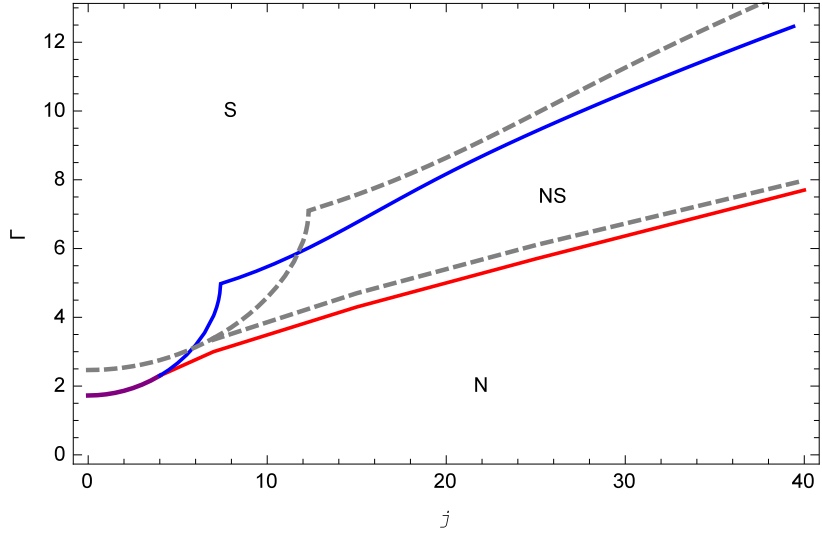}}
\caption{\label{dGSN}(Color online) Phase diagram for normal contacts with de Gennes parameter $b=0.2L$. S stands for the stationary regime and N for the normal regime.
The blue line is the stability limit when either the current or the temperature decreases, the red line is the stability limit when they increase, and the purple line is the stability limit in both directions. 
The dashed grey curves are the stability lines for $b=0$, presented in Fig.\ \ref{r0}.}
\end{figure}

Figure \ref{dGSN} is the phase diagram that we found for $b=0.2L$. We see that the boundaries move to higher temperatures and the Hopf singularity moves to a lower current density, but the topology is the same as in the case $b=0$, and the stationary regime still exists.

\begin{figure}
\scalebox{0.85}{\includegraphics{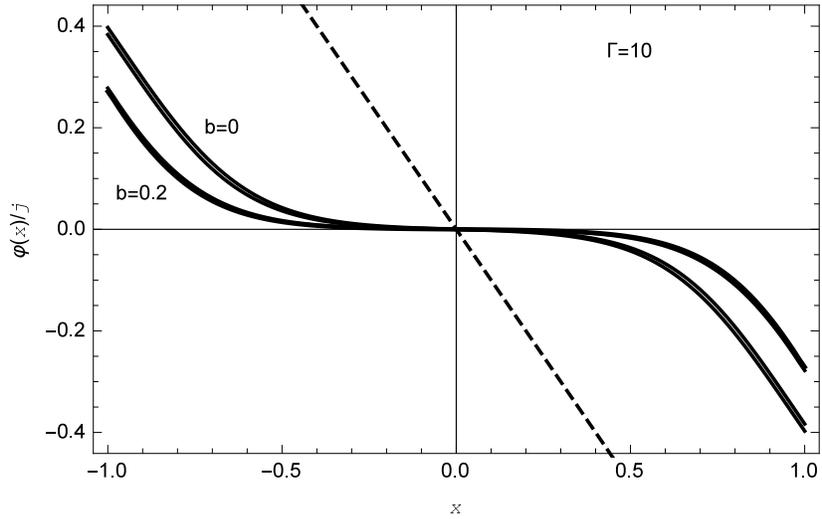}}
\caption{\label{Vstat} Potential along the wire in the stationary regime, divided by the current density, for $b=0$ and $b=0.2$, for $\Gamma =10$. The thickness of the lines covers the variation of $\varphi (x)/j$ within the range $0<j\le 40$. The dashed line describes $\varphi (x)/j$ for the normal regime.}
\end{figure}

Figure \ref{Vstat} shows the potential along the wire when it is in the stationary regime, for $b=0$ and for $b=0.2$, at a representative value of $\Gamma $. In the range $j\le 40$, $\varphi (x)$ is practically proportional to the current density.

\section{Conclusions}
We have studied the patterns of current flow along a one dimensional superconducting wire, for various boundary conditions, for current densities smaller than $160\sigma k_BT_c\xi^2(0)/\pi eL^3$ and temperatures above $T_c[1-12\xi^2(0)/L^2]$.

A possibility that was predicted long ago,\cite{Bara} but to our knowledge has never been detected, is the existence of a stationary regime, in which all observable fields are independent of time. The stationary regime might find applications in cases in which oscillations would induce disturbances (e.g., back action from a SQUID).

Our results clarify to what extent and in what sense the stationary regime can exist when the order parameter is not forced to vanish at the extremes of a superconducting wire. If the contacts are normal and the de Gennes boundary condition is assumed, the stationary regime still exists and the phase diagram is qualitatively unchanged, even if the de Gennes length is not negligible in comparison to the length of the wire.

If the contacts are superconducting, but weaker than the wire, the phase diagram changes qualitatively, and several new features arise. The role of the normal regime in Refs.\ \onlinecite{koby} and \onlinecite{Bara} is inherited by a periodic regime with one phase slip center, and the order parameter converges nonuniformly to zero as the current density increases. Part of the region in which Refs.\ \onlinecite{koby} and \onlinecite{Bara} found the stationary regime, becomes now fully superconducting. The fully superconducting regime and the periodic regime with one PSC are mediated by one or several periodic regimes with two PSC. In the limit that the contacts are made of very poor superconductors, the regime with two PSC becomes similar to the stationary state, since the PSC approach the extremes of the wire, and the time average of the potential felt by Cooper pairs is uniform between them.

In the regime with two PSC, the two phase slips may be simultaneous, but there may also be a time lag between them. When a time lag is present, the solution of the dynamic equations breaks the symmetry of the equations, and $|\psi (-x,t)|\neq |\psi (x,t)|$.

\begin{acknowledgments}
This research was supported by the Israel Science Foundation, grant No.\ 249/10. Numeric evaluations were performed using computer facilities of the Technion---Israel Institute of Technology. The author is indebted to John Kirtley, Jacob Rubinstein and Eli Zeldov for their answers to his enquiries.
\end{acknowledgments}

\appendix
\section{Numerical solution of Eqs.\ (\ref{tdgl})--(\ref{bcSC})}
The wire was discretized into a grid of $2^7$ vertices and time into steps of $3\times 10^{-5}t_0$. We started with an arbitrary function for the order parameter $\psi (x)$; if available, we took a function obtained from evolution with similar parameters, and otherwise we usually took $\psi\equiv r\Gamma^{1/2}$.

Derivatives with respect to $x$ were otained as follows: a fast Fourier transformation was performed on $\psi (x)$, and the Fourier transformed function was multiplied by $(ik)^p$, where $k$ is the reciprocal variable of $x$ and $p$ is the order of the desired derivative. We then transformed back to $x$-space and obtained $\partial^p\psi /\partial x^p$. 

The problem with this method is that it requires that $\psi$ be periodic with period $2L$, and in particular that $\psi (-L)=\psi (L)$. Since this is not necessarily the case, we start by evaluating the derivative of $\psi '(x)=\psi (x)-[\psi (L)-\psi (-L)]x/(2L)$, which does obey $\psi' (-L)=\psi' (L)$,
and subsequently add the derivative of $[\psi (L)-\psi (-L)]x/(2L)$. We are still left with the problem that the periodic extension of $\psi '$ is not smooth at $\pm L$ and we cannot evaluate its derivatives at the boundaries. However, $\psi_{xx}$ is not required at the boundaries, and $\psi_x(\pm L)$ enters the problem after multiplication by the length $L/2^6$ of a segment, so that $\psi_x(\pm L)$ can be evaluated as a finite difference.

The additional stages of the solution of Eqs.\ (\ref{tdgl})--(\ref{bcSC}) pose no numerical difficulty: $\varphi (x)$ can be obtained by trapezoidal integration of Eq.~(\ref{Ohm}) and the evolution of $\psi$ can be followed by Euler iteration. Likewise, the phases of $\psi (\pm L)$ can be updated by subtraction of the product of the time step times $\varphi (\pm L)$.


\begin{thebibliography}{99}
\bibitem{Ivlev}B. I. Ivlev and N. B. Kopnin, Adv. Phys. {\bf 33}, 47 (1984).
\bibitem{Tidecks}R. Tidecks, {\it Current-Induced Nonequilibrium Phenomena in Quasi-One-Dimensional Superconductors} (Springer, Berlin,
1990).
\bibitem{koby} J. Rubinstein, P. Sternberg, and Q. Ma, Phys. Rev. Lett. {\bf 99}, 167003 (2007).
\bibitem{Lydia} L. Peres-Hari, J. Rubinstein, and P. Sternberg, Physica D, {\bf 261}, 31 (2013).
\bibitem{Berdiy} G. Berdiyorov, K. Harrabi, F. Oktasendra, K. Gasmi, A. I. Mansour, J. P. Maneval, and F. M. Peeters, Phys. Rev. B {\bf 90}, 054506 (2014).
\bibitem{Almog} Y. Almog, L. Berlyand, D. Golovaty, and I. Shafrir, arXiv:1409.2128\,.
\bibitem{Bara} L. Kramer and A. Baratoff, Phys. Rev. Lett. {\bf 38}, 518 (1977).

\bibitem{shim} S. Kallush and J. Berger, Phys. Rev. B {\bf 89}, 214509 (2014).
\bibitem{Kopnin}N.B. Kopnin, {\it Theory of Nonequilibrium Superconductivity} (Clarendon Press, Oxford, 2001).
\bibitem{S}D.Y. Vodolazov, F.M. Peeters, L. Piraux, S. Ma´te´fi-Tempfli, and S. Michotte, Phys. Rev. Lett. {\bf 91}, 157001 (2003).
\bibitem{Lydia1} J. Kim, J. Rubinstein, and P. Sternsberg, Physica C {\bf 470}, 630 (2010). 
\bibitem{DJ} G. J. Dolan and L. D. Jackel, Phys. Rev. Lett. {\bf 39}, 1628 (1977).
\bibitem{dG} P. G. de Gennes, {\it Superconductivity of Metals and Alloys} (Westview Press, Boulder, 1999); Rev. Mod. Phys. {\bf 36}, 225 (1964).
\bibitem{Zum}J. Rubinstein, P. Sternberg, and K. Zumbrun, Arch. Rational Mech. Anal. {\bf 195}, 117 (2010).

\end{thebibliography}
\end{document}